# Bidding to the Top: VCG and Equilibria of Position-Based Auctions

Gagan Aggarwal    Jon Feldman    S. Muthukrishnan

August 7, 2018

Google[1]
{gagana,jonfeld,muthu}@google.com

**Abstract**

Many popular search engines run an auction to determine the placement of advertisements next to search results. Current auctions at Google and Yahoo! let advertisers specify a single amount as their bid in the auction. This bid is interpreted as the maximum amount the advertiser is willing to pay per click on its ad. When search queries arrive, the bids are used to rank the ads linearly on the search result page. The advertisers pay for each user who clicks on their ad, and the amount charged depends on the bids of all the advertisers participating in the auction. In order to be effective, advertisers seek to be as high on the list as their budget permits, subject to the market.

We study the problem of ranking ads and associated pricing mechanisms when the advertisers not only specify a bid, but additionally express their preference for positions in the list of ads. In particular, we study *prefix position auctions* where advertiser $i$ can specify that she is interested only in the top $\kappa_i$ positions.

We present a simple allocation and pricing mechanism that generalizes the desirable properties of current auctions that do not have position constraints. In addition, we show that our auction has an *envy-free* [4] or *symmetric* [8] Nash equilibrium with the same outcome in allocation and pricing as the well-known truthful Vickrey-Clarke-Groves (VCG) auction. Furthermore, we show that this equilibrium is the best such equilibrium for the advertisers in terms of the profit made by each advertiser. We also discuss other position-based auctions.

[1] Google, Inc., 1440 Broadway, New York, NY, 10018. Phone: 212-589-8831, fax 212-624-9605.

# 1 Introduction

In the sponsored search market on the web, advertisers bid on keywords that their target audience might be using in search queries. When a search query is made, an online (near-real time!) auction is conducted among those advertisers with matching keywords, and the outcome determines where the ads are placed and how much the advertisers pay. We will first review the existing auction model before describing the new model we study (a description can be found in Chapter 6 of [1]).

## 1.1 Current Auctions.

We describe a general auction model for sponsored search. Consider a specific query consisting of one or more *keywords*. When a user issues that search query, the search engine not only displays the results of the web search, but also a set of "sponsored links." In the case of Google, Yahoo, and MSN, these ads appear on a portion of the page near the right border, and are linearly ordered in a series of slots from top to bottom. (On Ask.com, they are ordered linearly on the top and bottom of the page).

Formally, for each search query, we have a set of $n$ advertisers interested in advertising on that query. This set is usually derived by taking a union over the sets of advertisers interested in the individual keywords that form the query. Advertiser $i$ bids $b_i$, which is the maximum amount the advertiser is willing to pay for a click. There are $k < n$ positions available for advertisements. When a query for that keyword occurs, an online auction determines the set of advertisements, their placement in the available positions, and the price per click each has to pay.

The most common auction mechanism in use today is the *generalized second-price* (GSP) auction (sometimes also referred to as the *next-price auction*). Here the ads are ranked in decreasing order of bid, and priced according to the bid of the next advertiser in the ranking. In other words, suppose wlog that $b_1 \geq b_2 \geq \cdots \geq b_n$; then the first $k$ ads are placed in the $k$ positions, and for all $i \in [1, k]$, bidder $i$ gets placed in position $i$ and pays $b_{i+1}$ per click.[2]

We note two properties ensured by this mechanism:

1. *(Ordering Property)* The ads that appear on the page are ranked in decreasing order of $b_i$.

2. *(Minimum Pay Property)* If a user clicks on the ad at position $i$, the advertiser pays the minimum amount she would have needed to *bid* in order to be assigned the position she occupies.

Search engine companies have made a highly successful business out of these auctions. In part, the properties above have dissuaded advertisers from trying to game the auction. In particular, the minimum-pay property ensures that an advertiser has no incentive to lower a winning bid by a small amount in order to pay a lower price for the same position. Still, the GSP auction is not truth-revealing, that is, an advertiser may be incentivized to bid differently than her true value under certain conditions [2].

Only recently have we obtained a detailed formal understanding of the properties of this auction. Authors in [4, 8, 2] have analyzed the auction in terms of its equilibria. They show that when the click-through rates are *separable*, i.e. the click-through rate of an ad at a given position is the product of an ad-specific factor and a position-specific factor, the GSP has a Nash equilibrium whose outcome is equivalent to the famous Vickrey-Clarke-Groves (VCG) mechanism [9, 3, 5] which is known to be truthful. [4, 8] go on to show that this equilibrium is *envy-free*, that is, each advertiser prefers the current outcome (as it applies to her) to being placed in another position and paying the price-per-click being paid by the current occupant of the position. Further, among all the envy-free equilibria, the VCG equilibrium is bidder-optimal; that is, for each advertiser, her price-per-click is minimum under this equilibrium. We note that when the click-through rates are separable, the outcome produced by the VCG mechanism has the ordering property. Authors in [2] also show that even when the click-through rates are arbitrary, there is a pricing method with the ordering property that is truthful. (This pricing method reduces to the VCG pricing method when the click-through rates are separable.) Furthermore, they show that the GSP has a Nash equilibrium that has the same outcome as their mechanism. Together, these results provide some understanding of the current auctions. That in turn provides confidence in the rules of the auction, and helps support the vast market for search keyword advertising.

---
[2]The Google auction actually ranks according to $w_i b_i$, for some weight $w_i$ related to the quality of the ad, and then sets the price for bidder $i$ to $w_{i+1} b_{i+1} / w_i$. All our results generalize to this "weighted" bid case as well.

**1.2 Emerging Position-based Auctions.** As this market matures, advertisers are becoming increasingly sophisticated. For example they are interested in the relative performance of their ads and keywords, and so the search engines provide tools to track keyword-level statistics. As advertisers learn when and where their ads are most effective, they need more control over their campaigns than is provided by simply stating a keyword and a bid.

One of the most important parameters affecting the performance of an advertisement is its position on the page. Indeed, the reason the auction places the ads in descending order on the page is that the higher ads tend to get clicked on more often than the lower ones. In fact, having an ad place higher on the page not only increases the chances of a click, it also has value as a branding tool, regardless of whether the ad gets clicked. Indeed, a recent empirical study by the Interactive Advertising Bureau and Nielsen//NetRatings concluded that higher ad positions in paid search have a significant brand awareness effect [7]. Because of this, advertisers would like direct control over the position of their ad, beyond just increasing the bid. Ideally, the search engine would conduct a more general auction that would take such position preferences into account; we refer to this as a *position-based auction*.

**1.3 Our Results.** In this paper, we initiate the study of position-based auctions where advertisers can impose position constraints.

In particular, we study the most relevant case of *prefix* position constraints, inspired by the branding advertiser: advertiser $i$ specifies a position $\kappa_i$ and a bid $b_i$, which says that the advertiser would like to appear only in the top $\kappa_i$ positions (or not at all) and is willing to pay at most $b_i$ per click. Upon receiving bids from a set of $n$ such advertisers, the search engine must conduct an auction and place ads into $k$ positions while respecting the prefix constraints.

Our main results are as follows. We present a simple auction mechanism that has both the ordering and the minimum pay property, just like the current auctions. The mechanism is highly efficient to implement, taking near-linear time. Further, we provide a characterization of its equilibria. We prove that this auction has a Nash equilibrium whose outcome is equivalent in allocation and pricing to that of VCG. Additionally, we prove that this equilibrium is *envy-free* and that among all envy-free equilibria, this particular one is bidder-optimal.

Our results generalize those in [4, 8], which proved the same thing for the GSP without any position constraints. The main difficulty in generalizing these results lies in the fact that once you allow position constraints, the allocation function of VCG no longer obeys the ordering property, thus making it challenging to engineer an equilibrium whose outcome is identical. The main technical contributions of the paper are new structural properties of the VCG allocation that allow us to relate the VCG allocation with an auction that preserves the ranking order.

In the future, advertisers may want even more fine-grained control over ad position. We discuss more general position-based auctions at the end of the paper.

**Outline.** In Section 2, we present a few natural mechanisms for prefix position auctions and show their limitations before describing our proposed "top-down auction" which has both the ordering and the minimum-pay property. In Section 3, we analyze the equilibrium properties of our top-down auction. Concluding remarks, including discussions of more general position-bidding problems, are in Section 4.

## 2 Prefix Position Auction Mechanisms

Formally, the prefix position auction problem is as follows. There are $n$ advertisers for a search keyword. They submit bids $b_1, \ldots, b_n$ respectively. There are $k$ positions for advertisements numbered $1, \ldots, k$, top to bottom. Each advertiser $i \in \{1, \ldots, n\}$ also submits a cutoff position $\kappa_i \leq k$, and requires that their advertisements should not appear below position $\kappa_i$.

An auction mechanism consists of two functions:

- an *allocation* function that maps bids to a matching of advertisers to positions, as well as

- a *pricing* function that assigns a price per click $\text{ppc}_j$ to each position won by an advertiser. We restrict our attention to mechanisms where the prices always respect the bids; i.e., we have $\text{ppc}_j \leq b_i$ if $i$ is assigned to $j$.

A natural allocation strategy that retains the ordering property is as follows: rank the advertisers in decreasing order of $b_i$ as in GSP. Now, go through the ranking one position at a time, starting at the top; if you encounter an advertiser that appears below her bottom position $\kappa_i$, remove her from the ranking and move everyone below that position up one position, and continue checking down the rest of the list.

Two natural pricing strategies immediately come to mind here:

1. Set prices according to the subsequent advertiser in the ranking *before* any advertiser is removed, or

2. Set prices according to the subsequent advertiser in the ranking *after* all the advertisers are removed (more precisely, the ones that appear below her position).

It turns out that neither of these options achieves the minimum pay property as shown by the following examples. Assume for the sake of these examples that $0.05 is the amount used to charge for the last position.

1. Suppose we set prices before removing out-of-position advertisers. Now suppose we have the following ranges and bids where the number in parentheses is the position constraint $\kappa_i$:

    A: (5) $5    B: (5) $4    C: (5) $3    D: (2) $2    E: (5) $1

We run the auction, and the order is (A, B, C, D, E). If we price now, the prices are ($4, $3, $2, $1, $0.05). Bidder D gets removed and so we end up with (A, B, C, E), and we charge ($4, $3, $2, $0.05). However, if bidder C had bid $1.50, which is below what she was charged, the auction would still have ended up as (A, B, C, E). Thus, the minimum pay property is violated because we are charging too much.

For a more intuitive reason why this is a bad mechanism, it would allow a form of "ad spam". Suppose a bidder sets her bottom cutoff to (2), but then bids an amount that would never win position one or two. In this case, she drives up the price for those that are later in the auction (e.g., competitors), at no risk and no cost to herself.

2. Now suppose we set the prices after removing out-of-position advertisers, and we have the following bids and prefix constraints:

    A: (5) $5     B: (5) $4     C: (2) $3     D: (5) $2

    We run the auction, and the order is (A, B, C, D). Now we remove C and we get the order (A, B, D). We price according to this order and so the prices are ($4, $2, $0.05). Bidder B bid $4 and paid $2; however, if B changed her bid to $2.50, then bidder C would have gotten second position. Thus the minimum pay property is violated, but this time because we are charging too little.

    As for intuition, this option opens up a possible "race for the bottom" situation. Suppose we have a group of bidders only interested in positions 1-4 (perhaps because those appear on the page without scrolling). The winners of the top three positions pay according to the competitive price for those top positions, but the winner of position 4 pays according to the winner of position 5, who could be bidding a much lower amount. Thus, these top bidders have an incentive to lower their prices so that they can take advantage of this bargain.

But now consider a third alternative, which will turn out to be the one that achieves the minimum-pay property: *For each advertiser that is allocated a particular position j, set the price according to the first advertiser that appears later in the original ranking that included j in her range.*

For an example of this pricing method, consider the situations from (1),(2) above:

1. A: (5) $5     B: (5) $4     C: (5) $3     D: (2) $2     E: (5) $1

   The advertisers are ranked (A, B, C, D, E), and then (A, B, C, E) after removal. The price for A is set to $4, since B had position 1 in its range. Similarly, the price for B is set to $3 since C had position 2 in its range. The price for C is set to $1, however, since D did not include position 3 in its range. The price for C is set to $0.05.

2. A: (5) $5     B: (5) $4     C: (2) $3     D: (5) $2

   The advertisers are ranked (A, B, C, D) and after removal we get (A, B, D). The price for A is $4, but the price for B is now $3; even though C did not win any position, it was still a participant in the auction, and was bidding for position 2. The price for D is $0.05.

**2.1 Top-down Auction.** The preceding examples demonstrate that one needs to be careful in setting prices when position constraints are involved. Here we define an auction mechanism for prefix position constraints that is equivalent to the final proposal above, and is easily seen to have the minimum-pay property:

**Definition 1** *The* top-down auction mechanism *works as follows: For each position in order from the top, iteratively run a simple second-price auction (with one winner) among those advertisers whose prefix range includes the position being considered. (We use $b_i$ to determine the ranking.) By a "simple second-price auction," we mean that the highest bidder in the auction is allocated the position, and pays a price-per-click equal to the second-highest bid. This winner is then removed from the pool of advertisers for subsequent auctions and the iteration proceeds.*

This mechanism is exceedingly easy to implement, taking time $O(n \log n)$.

## 3 Analysis of the top-down prefix auction

Now that we have found a natural generalization of the GSP to use with prefix position constraints, we would like to know what properties this auction has. Since the GSP is a special case, we already know that the auction is not truthful [2]. But from [8, 4, 2] we at least know something about the equilibria of the GSP auction. It is natural to ask whether or not these results hold true in our more general setting.

In this section, we answer this in the affirmative, and prove that the top-down prefix auction has an "envy-free" Nash equilibrium whose outcome (in terms of allocation and pricing) is equivalent to that of VCG. ("Envy-freeness" is a stronger condition than is imposed by the Nash equilibrium, dictating that no bidder envies the allocation and price of any other bidder.) We go on to prove that this equilibrium is bidder-optimal envy-free Nash equilibrium in the sense that it maximizes the "utility" (or profit) made by each advertiser.

We now define the "utility" functions that each advertiser seeks to maximize.

Each position $j$ has an associated *click-through rate* $c_j > 0$ which is the probability that a user will click on an ad in that position. Using the idea that higher positions receive more clicks, we may assume $c_1 > c_2 > \cdots > c_k$. To make the discussion easier, we will abuse this notation and say that an ad in position $j$ "receives $c_j$ clicks," even allowing $c_j > 1$ for some examples.

Each advertiser has a valuation $v_i$ it places on a click, as long as that click comes from one of its desired positions. Using the "branding" motivation, we assume a valuation of $-\infty$ if an ad even *appears* at a position below its bottom cutoff $\kappa_i$. Since $c_j > 0$ for all positions $j$, we can (equivalently) think of this as a valuation of $-\infty$ on a *click* below position $\kappa_j$. So, given some total price $p$ (for all the clicks) for a position $j$, the utility of bidder $i$ is $u_i = c_j v_i - p$ if $j \leq \kappa_i$, and $-\infty$ otherwise.

Note that we are making the assumption that click-through rates are dependent only on the position and not on the ad itself. Our results hold as long as the click-through rates are *separable*, i.e. the click-through rate of an ad at a given position is the product of a per-position factor and a per-advertiser factor. More general forms of click-through rate would require further investigation.

**3.1 The Vickrey-Clarke-Groves (VCG) Auction.** The Vickrey-Clarke-Groves (VCG) auction mechanism [9, 3, 5] is a very general technique that can be applied in a wide range of settings. Here we give its application to our problem. For a more general treatment, we refer the reader to Chapter 23 of [6].

Let OPT represent the allocation of bidders to positions maximizes the *total valuation* on the page; i.e., OPT is a matching $M$ of advertisers $i$ to positions $j$ that respects the position constraints ($j \leq \kappa_i$), and maximizes $\sum_{(i,j) \in M} v_i c_j$. Note that this assignment could also have empty slots, but they must be contiguous at the bottom end.

The OPT allocation is the most "efficient" allocation, but an allocation function in an auction mechanism has access to the bids $b_i$ not the valuations $v_i$. So instead, the VCG allocation $M^*$ is the matching $M$ that maximizes $\sum_{(i,j) \in M} b_i c_j$.

Intuitively, the VCG price for a particular bidder is the total difference in others' valuation caused by that bidder's presence. To define this pricing function formally, we need another definition: Let $M^*_{-x}$ be the VCG allocation that would result if bidder $x$ did not exist. More formally, this allocation is the matching $M$ that does not include bidder $x$ and maximizes $\sum_{(i,j) \in M} b_i c_j$.

The VCG price for bidder $i$ in position $j$ is then $p_i = M^*_{-i} - M^* + c_j b_i$. (Here we are abusing notation and using $M^*$ and $M^*_{-i}$ to denote the total valuation of the allocation as well as the allocation itself.) Note that $p_j$ is a *total* price for all clicks at that position, not a per-click price.

Only in the case that $b_i = v_i$ does the VCG mechanism actually successfully compute OPT. However, it is well-known (see [6] for example) that the pricing method of VCG ensures that each bidder is incentivized to actually reveal their true valuation and set $b_i = v_i$. This holds *regardless of the actions of the other bidders,* a very strong property referred to as "dominant-strategy truthfulness."

Thus in equilibrium, we get $b_i = v_i$, and $M^* =$ OPT. We also have $M^*_{-i} =$ OPT$_{-i}$, where OPT$_{-i}$ is the OPT allocation that would result if bidder $i$ did not exist.

For convenience, for the remainder of paper we rename the bidders by the slots to which they were assigned in OPT, even when we are talking about the top-down prefix auction. The unassigned bidders are renamed to $(k+1, \ldots, n)$ arbitrarily. We will use $p_i =$ OPT$_{-i} -$ OPT $+ c_i v_i$ to denote the VCG price (at equilibrium) for position (and bidder) $i$.

**3.2 Envy-Free Nash Equilibria and the GSP Auction.** The VCG mechanism is desirable because it has an equilibrium that results in the most efficient allocation according to the true valuations of the bidders. Furthermore this equilibrium occurs when each bidder actually reveals their true valuations. The GSP auction (without position constraints) does not have this second property, but in fact it does have the first: namely that it has an equilibrium whose allocation is the most efficient one (this was proved in [4, 8, 2]). Furthermore, this equilibrium also results in the same *prices* that result from VCG. This somehow validates the GSP from an incentive-compatibility point of view, and shows that the ordering property does not preclude efficiency.

This equilibrium also has the following property:

**Definition 2** *An allocation and pricing is an* envy-free equilibrium *if each bidder prefers the current outcome (as it applies to her) to being placed in another position and paying the price-per-click being paid by the current occupant of the position.*

Moreover, among all envy-free Nash equilibria, this particular one is *bidder-optimal*, in the sense that it results in the lowest possible price for each particular advertiser.

**3.3 Equilibrium in the top-down auction.** It is natural to ask if all these properties also hold true in the presence of position constraints. One of the difficulties in proving this comes from the fact that the VCG allocation no longer preserves the ranking order, as shown by the following simple example. Suppose advertiser A has bottom cutoff (2) and a bid of $2, advertiser B has cutoff (1) and a bid of $1, and we have $c_1 = 101$ and $c_2 = 100$. The VCG allocation gives position 1 to B and position 2 to A, for a total revenue of $\approx \$300$. The top-down auction will give position 1 to A and position 2 will be unfilled. The revenue is equal to $\approx \$200$.

Despite this, it turns out that there is an equilibrium of the top-down auction where bidders end up in the optimal allocation, which we prove in our main theorem:

**Theorem 1** *In the top-down prefix auction, there exists a set of bids and stated position cutoffs such that*

(a) *each bidder is allocated to the same slot as she would be in the dominant-strategy equilibrium of VCG,*

(b) *the winner of each slot pays the same total price as she would have in the dominant-strategy equilibrium of VCG, and*

(c) *the bidders are in an envy-free Nash equilibrium.*

*Furthermore (d), for each advertiser, her utility under VCG outcomes is the maximum utility she can make under any envy-free equilibrium. In other words, a VCG outcome is a bidder-optimal envy-free equilibrium of the top-down auction.*

The remainder of this section is devoted to proving this theorem. The bids that will satisfy this theorem are in fact quite simple: we set $b_i = p_{i-1}/c_{i-1}$ for all bidders $i$ assigned in OPT. Thus, if we could prove that $b_1 > b_2 > \cdots > b_k$, we would get that the top-down auction assigns the bidders exactly like OPT and sets the same prices (modulo some details about the last bidder and those not assigned in OPT, which we discuss later). This would prove (a) and (b) above.

To show that the bids are indeed decreasing, and to show (c), it turns out that we need to prove some technical lemmas about the difference between OPT and $\text{OPT}_{-i}$ for some arbitrary bidder $i$. In $\text{OPT}_{-i}$, some bidder $i'$ takes the place of $i$ (unless $i$ is in the last slot, in which case perhaps no bidder takes this slot). In turn, some bidder $i''$ takes the slot vacated by $i'$, etc., until either the vacated slot is the bottom slot $k$, or some previously unassigned bidder is introduced into the solution. We call this sequence of bidder movements ending at slot $i$ the "chain" of moves of $\text{OPT}_{-i}$. Note that the chain has the property that it begins either with an unassigned bidder, or with the bidder from the last slot and ends at slot $i$.

If we consider the slots not on the chain, we claim that (wlog) the assignment does not change on these slots when we go from OPT to $\text{OPT}_{-i}$. This is easily seen by substituting a purported better assignment on these slots back into OPT. Note that this implies that $\text{OPT}_{-i}$ has at most one new bidder (that wasn't in OPT), and that no bidder besides $i$ that was assigned in OPT has dropped out. We chain is said to have *minimum length* if there is no shorter chain that achieves the same valuation as $\text{OPT}_{-i}$.

A *link* in this chain refers to the movement of a bidder $i$ from slot $i$ to some slot $i'$. We say that this is a *downward* link if $i' > i$; otherwise it is an upward link.

**Lemma 1** *The minimum length chain for $OPT_{-i}$ does not contain a downward link followed by an upward link.*

**Proof:** Suppose it does contain such a sequence. Then, some bidder $i_1$ moved from slot $i_1$ to slot $i_2 > i_1$, and bidder $i_2$ moved from slot $i_2$ to a slot $i_3 < i_2$. An alternate solution, and thus a candidate solution for $OPT_{-i}$ is to have bidder $i_1$ move from slot $i_1$ to slot $i_3$, have bidder $i_2$ remain in slot $i_2$, and keep everything else the same. (Bidder $i_1$ can move to slot $i_3$ since $i_3 < i_2$ and $i_2$ is in range for bidder $i_1$ (by the fact that $i_1$ moved to $i_2$ in $OPT_{-i}$).)

The difference between the two solutions is $c_{i_2}(v_{i_2} - v_{i_1}) + c_{i_3}(v_{i_1} - v_{i_2}) = (c_{i_3} - c_{i_2})(v_{i_1} - v_{i_2})$. We know that $c_{i_3} > c_{i_2}$ since $i_3 < i_2$. We also know that $v_{i_1} \geq v_{i_2}$ since otherwise OPT would have switched bidders $i_1$ and $i_2$ (note again that bidder $i_1$ can move to slot $i_2$, since it did so in $OPT_{-i}$). Thus the difference is non-negative, and so this alternate solution to $OPT_{-i}$ has either greater valuation or a shorter chain. ∎

**Lemma 2** *Let $x$ and $y$ be arbitrary bidders assigned to slots $x$ and $y$ in OPT, where $x < y$. Then, (i) if slot $y$ is in the range of bidder $x$, we have $OPT_{-y} \geq OPT_{-x} + c_y(v_x - v_y)$, and (ii) $OPT_{-x} \geq OPT_{-y} + c_x(v_y - v_x)$.*

**Proof: (i)** Consider the assignment of bidder $y$ in $OPT_{-x}$. Recall that for any $i$, all bidders besides $i$ present in OPT are also present in $OPT_{-i}$. Thus $y$ is present somewhere in $OPT_{-x}$. Note also that the chain for $OPT_{-x}$ ends at slot $x$, and so if $y$ is present in this chain, it cannot be on a downward link; otherwise the chain would have to contain a downward link followed by an upward link (contradicting Lemma 1), since $x$ is above $y$. Thus we may conclude that $y$ ends up in position $y' \leq y$.

Since slot $y$ is in range for bidder $x$ by assumption, we also have that $y'$ is in range for bidder $x$; thus we can construct a candidate solution for $OPT_{-y}$ by replacing (in $OPT_{-x}$) bidder $y$ with bidder $x$. We may conclude that

$$OPT_{-y} \geq OPT_{-x} + c_{y'}(v_x - v_y)$$

Since $y' \leq y$ we have $c_{y'} \geq c_y$ and so we get (i).

**(ii)** This time we need to consider the assignment of $x$ in $OPT_{-y}$. By the same logic as above, bidder $x$ is present somewhere, and if $x$ either stayed in the same place of moved up, we can replace $x$ with $y$ (in $OPT_{-y}$) to get a candidate for $OPT_{-x}$, and we are done. The only remaining case is when $x$ moves down in $OPT_{-y}$ and this is a bit more involved.

Consider the section of the chain of $OPT_{-y}$ from bidder $x$ to the end at bidder $y$ (who is below $x$). Since $x$ is on a downward link, and downward links cannot be followed by an upward link (Lemma 1), it must be the case that this section of the chain is entirely downward links. Let $x \to x_1 \to x_2 \to \cdots \to x_\ell \to y$ be this chain, and so we have $x < x_1 < x_2 \cdots < x_\ell < y$.

We write the assignment of OPT to these $\ell + 2$ places using the notation $[x, x_1, x_2, \ldots, x_\ell, y]$, and consider other assignments to these slots using the same notation. The solution $OPT_{-y}$ assigns these slots as $[w, x, x_1, \ldots, x_\ell]$, where $w$ is the bidder before $x$ in the chain. For notational purposes define $x_{\ell+1} = y$.

Consider the following alternate solution constructed from $OPT_{-y}$: change only the assignments to these special $\ell + 2$ slots to $[w, x_1, \ldots, x_\ell, y]$. This is a candidate for $OPT_{-x}$ and so by calculating the difference in valuation between this candidate solution and $\text{OPT}_{-y}$ we get

$$\text{OPT}_{-x} \geq \text{OPT}_{-y} + \left(\sum_i^\ell v_{x_i}(c_{x_i} - c_{x_{i+1}})\right) + c_y v_y - c_{x_1} v_x \tag{1}$$

Putting this aside for now, consider the following alternate solution for OPT. Take the assignment in OPT and change the assignment to only those $\ell + 2$ positions to $[y, x, x_1, \ldots, x_\ell]$. This is feasible since $y$ moves up, and the remaining changes are identical to $\text{OPT}_{-y}$. Since this solution must have valuation at most that of OPT, we get that

$$c_x v_y + c_{x_1} v_x + \sum_1^\ell v_{x_i} c_{x_{i+1}} \leq c_x v_x + \left(\sum_1^\ell v_{x_i} c_{x_i}\right) + c_y v_y$$

$$\iff c_x(v_y - v_x) \leq \left(\sum_1^\ell v_{x_i}(c_{x_i} - c_{x_{i+1}})\right) + c_y v_y - c_{x_1} v_x$$

This, combined with Equation (1), implies (ii). ∎

Now we are ready to prove the first part of our main theorem, namely that our bids give the same outcome as VCG, and that they are indeed an envy-free equilibrium.

**Proof of Theorem 1(a-c)** The bids of the equilibrium are defined as follows. For all bidders $i > 1$ assigned in OPT, we set $b_i = p_{i-1}/c_{i-1}$. We set $b_1$ to any number greater than $b_2$. For all bidders assigned in OPT, we set their stated cutoff to their true cutoff $\kappa_i$. If there are more than $k$ bidders, then for some bidder $\alpha$ that was not assigned in OPT, we set $b_\alpha = p_k/c_k$, and set the stated cutoff of bidder $j$ to the bottom slot $k$. For all other bidders not in OPT, we set their bid to zero, and their cutoff to their true cutoff.

Consider two arbitrary bidders $x$ and $y$ assigned in OPT, where $x < y$. Using Lemma 2(ii), we get $\text{OPT}_{-x} \geq \text{OPT}_{-y} + c_x(v_y - v_x)$. Substituting for $\text{OPT}_{-x}$ and $\text{OPT}_{-y}$ using the definitions of $p_x$ and $p_y$, respectively, we get:

$$\text{OPT} - c_x v_x + p_x \geq \text{OPT} - c_y v_y + p_y + c_x(v_y - v_x)$$

$$\iff \left(v_y - \frac{p_y}{c_y}\right) c_y \geq \left(v_y - \frac{p_x}{c_x}\right) c_x$$

Since $c_y < c_x$, we get $\frac{p_x}{c_x} > \frac{p_y}{c_y}$.

Since we chose $x$ and $y$ arbitrarily, we have just showed that $b_2 > \cdots > b_k$, and $b_k > b_\alpha$ if bidder $\alpha$ exists. We have $b_1 > b_2$ by definition, and all other bids are equal to zero. Thus the bids are decreasing in the VCG order, and so the Google auction will choose the same allocation as VCG. By construction, the Google auction will also have the same prices as VCG.

It remains to show that this allocation and pricing is an envy-free equilibrium. Consider again two bidders $x$ and $y$ assigned in OPT with $x < y$. The utilities of $x$ and $y$ are $u_x = c_x v_x - p_x$ and $u_y = c_y v_y - p_y$. We must show that $x$ does not envy $y$, and that $y$ does not envy $x$.

If $y$ is out of range of bidder $x$, then certainly $x$ does not envy $y$. If $x$ is in range of bidder $y$, then by Lemma 2(i), we get $\text{OPT}_{-y} \geq \text{OPT}_{-x} + c_y(v_x - v_y)$. Substituting for $\text{OPT}_{-y}$ and $\text{OPT}_{-x}$ using the definitions of $p_y$ and $p_x$, we get

$$\text{OPT} + p_y - c_y v_y \geq \text{OPT} + p_x - c_x v_x + c_y(v_x - v_y)$$

$$\iff c_y v_x - p_y \leq c_x v_x - p_x = u_x$$

This implies that $x$ does not envy $y$. Similarly, Lemma 2(ii) shows that $y$ does not envy $x$.

Now consider some bidder $z$ not assigned in OPT. We must show that bidder $z$ does not envy any bidder that is assigned a slot in the desired range of $z$. Consider some such bidder $y$; replacing $y$ with $z$ creates a candidate for $\text{OPT}_{-y}$. Thus we have $\text{OPT}_{-y} \geq \text{OPT} + c_y(v_z - v_y)$, which becomes $p_y = \text{OPT}_{-y} - \text{OPT} + c_y v_y \geq c_y v_z$. This implies that $z$ does not envy $y$. ∎

Now it remains to show the second part of Theorem 1, namely that among all envy-free equilibria, the one we define is optimal for each bidder. First we give a lemma showing that envy-freeness in the top-down auction implies that the allocation is the same as VCG. Then we use this to compare our equilibrium with an arbitrary envy-free equilibrium.

**Lemma 3** *Any envy-free equilibrium of the top-down auction has an allocation with optimal valuation.*

**Proof:** For the purposes of this proof, we will extend any allocation of bidders to slots to place all $n$ bidders into "slots". For this, we will introduce dummy slots indexed by integers greater than $k$, with click-through rate $c_i = 0$. We index the bidders according to their (extended) allocation in OPT.

For the purposes of deriving a contradiction, let $E, p$ be the allocation and pricing for an envy-free equilibrium of the Google auction such that the valuation of $E$ is less than OPT. Thus, $p_i$ refers to the price of slot $i$ in this envy-free equilibrium. Define a graph on $n$ nodes, one for each slot. For each bidder $i$, make an edge from $i$ to $j$, where $j$ is the slot in which bidder $i$ is placed in $E$; i.e., bidder $i$ is in slot $i$ in OPT and in slot $j$ in $E$. Note that this graph is a collection of cycles (a self-loop is possible, and is defined as a cycle).

Define the weight of an edge $(i, j)$ to be the change in valuation caused by bidder $i$ moving from slot $i$ in OPT to slot $j$ in $E$. So, we have that the weight of $(i, j)$ is equal to $v_i(c_j - c_i)$. Since the total change in valuation from OPT to $E$ is negative by definition, the sum of the weights of the edges is negative. This implies that there is a negative-weight cycle $Y$ in the graph, and so we have

$$\sum_{(i,j) \in Y} v_i(c_j - c_i) < 0. \tag{2}$$

By the fact that $E$ is envy-free, for each edge $(i, j)$, we also have that bidder $i$ would rather be in slot $j$ than in slot $i$ (under the prices $p$ imposed by the envy-free equilibrium). In other words, $v_i c_j - p_j \geq v_i c_i - p_i$. Rearranging and summing over the edges in $Y$, we get

$$\sum_{(i,j) \in Y} v_i(c_j - c_i) \geq \sum_{(i,j) \in Y} p_j - p_i = 0. \tag{3}$$

(The sum on the right-hand side equals zero from the fact that $Y$ is a cycle.) Equations (2) and (3) together give us a contradiction. ∎

Note that the profit of an advertiser $i$ is the same under all VCG outcomes, and is equal to the difference in valuation between OPT and $\text{OPT}_{-i}$.

**Proof of Theorem 1(d)** Consider some envy-free equilibrium $E$ of the Google auction. This equilibrium must have an allocation with optimal valuation (by Lemma 3). We will call this allocation OPT. Let $\{p_i^E\}_i$ be the price of slot $i$ in this equilibrium; We will rename the bidders such that bidder $i$ is assigned to slot $i$ by allocation OPT. Consider one such bidder $x$ assigned to slot $x$. Consider the chain $x_\ell \to x_{\ell-1} \to \cdots \to x_0 = x$ for $\text{OPT}_{-x}$. (Here bidder $x_j$ moves from slot $x_j$ in OPT to slot $x_{j-1}$ in $\text{OPT}_{-x}$.) By the fact that $E$ is envy-free, for all $j \in [0, \ell-1]$ we have

$$v_{x_{j+1}} c_{x_{j+1}} - p^E_{x_{j+1}} \geq v_{x_{j+1}} c_{x_j} - p^E_{x_j}$$
$$\iff \quad p^E_{x_j} \geq v_{x_{j+1}}(c_{x_j} - c_{x_{j+1}}) + p^E_{x_{j+1}}$$

(Each move is this chain is feasible, since it was made by $\text{OPT}_{-x}$.) Composing these equations for $j = 0, \ldots, \ell-1$, we get

$$p^E_x = p^E_{x_0} \geq v_{x_1}(c_{x_0} - c_{x_1}) + v_{x_2}(c_{x_1} - c_{x_2}) + \cdots + v_{x_\ell}(c_{x_{\ell-1}} - c_{x_\ell})$$

But note that each term of the right-hand side of this inequality represents the difference in valuation for a bidder on the chain of $\text{OPT}_{-x}$. Thus the sum of these terms is exactly the VCG price $p_x$, and we have $p^E_x \geq p_x$. Hence, the profit of advertiser $x$ under equilibrium $E$ is no less than her profit under VCG. ∎

## 4 Concluding Remarks

The generalized second-price auction has worked extraordinarily well for search engine advertising. We believe that the essential properties of this auction that make it a success are that it preserves the ranking order inherent in the positions, and that it is stable in the sense that no bidder has an incentive to change her bid by a small amount for a small advantage. We have given a simple new prefix position auction mechanism that preserves these properties and has the same equilibrium properties as the regular GSP.

A natural question arises if advertisers will have preference for positions that go beyond the top $\kappa_i$'s. It is possible that there are other considerations that make lower positions more desirable. For example, the last position may be preferable to the next-last. Also, appearing consistently at the same position may be desirable for some. Some of the advertisers may not seek the topmost positions in order to weed out clickers who do not persist through the topmost advertisements to choose the most appropriate one. Thus, there are a variety of factors that govern the position preference of an advertiser. In the future, this may lead to more general position auctions than the prefix auctions we have studied here. We briefly comment on two variants.

**Arbitrary Ranges.** If we allow top cutoffs (i.e., bidder $i$ can set a valuation $\alpha_i$ and never appear above position $\alpha_i$), we can consider running essentially the same top-down auction: For each position in order from the top, run a simple Vickrey auction (with one winner) among those advertisers whose range includes the position being considered; the winner is allocated the position, pays according to the next-ranked advertiser, and is removed from the pool of advertisers for subsequent auctions.

The difference here is that we can encounter a position $j$ where there are not advertisers willing to take position $j$, but there are still advertisers willing to take positions lower than $j$. (This cannot occur with prefix ranges.) On a typical search page, the search engine must fill in something at position $j$, or else the subsequent positions do not really make sense. Practically speaking, one could fill in this position with some sort of "filler" ad. Given some sort of resolution of this issue, the top-down auction maintains the minimum pay property for general ranges, by essentially the same argument as the prefix case in this paper.

However, the property that there is an equilibrium that matches the VCG outcome is no longer true, as shown by the following example:

**Example.** Suppose we have three bidders, and their ranges and valuations are given as follows: A (1,1) $3; B (2,3) $2; C (1,3) $1. We also have three positions, and we get 100, 99 and 98 clicks in them, respectively. The VCG outcome is an allocation of [A,B,C], and prices [$2, $1, $0] (for *all* clicks). To achieve this outcome in the Google range auction, we must have A with the highest bid, and it is $\infty$ wlog. Since C is the only other bidder competing for the first slot, the price of A (which must be $2) is determined by the bid of B, and thus $2 = p_1 = b_B c_1 = 100 b_B$. Therefore we have that $b_B = 2/100$. Since bidder C wins the second slot, we must have bidder C outbidding bidder B and so $b_C \geq b_B = 2/100$. The price of the second slot is also determined by the bid of B, and we get $p_2 = b_B c_2 = 99(2/100)$. This is inconsistent with the VCG price of $1.

**General Position Bids** One is tempted to generalize the position-based auction so that instead of enforcing a ranking, each advertiser submits separate bids for each position and the market decides which positions are better. Suppose we allowed such bids, and let $b_{i,j}$ denote the bid of advertiser $i$ for position $j$.

In this setting, the ranking property no longer makes sense, but it still might be interesting to consider the minimum-pay property. We do need to clarify our definition of this property; because the advertiser has control of more than just the bid that gave her the victory; we need to make sure that altering the *other* bids cannot give the advertiser a bargain for this particular position.

A natural mechanism and pricing scheme is as follows: Given the bids $b_{i,j}$, compute the maximum matching of bids to positions (i.e., the VCG allocation). Now for each winning bid $b^i_j$, do the following. delete all other edges from i, and lower this bid until the max matching no longer assigns $i$ to $j$. Set the price per click $\text{ppc}_j$ to the bid where this happens.

Note that this price has the property that if the winner of a position bids between the bid and the price, then they either get the same position at the same price, or perhaps one of their other bids causes them to get a different position. But, we still have the property that the winner cannot get the position she won for a lower price.

It turns out that this is exactly the VCG mechanism, as seen by the following argument. In the following, let $M$ be the valuation of the maximum matching, and for some $i$ let $M_{-i}$ be the valuation of the maximum matching that does not include bidder $i$. Note also that if bidder $i$ is assigned position $j$, the VCG price is $M_{-i} - (M - b_{i,j} c_j)$.

In the suggested auction, when setting the price for bidder $i$, consider the moment when the bid is lowered to $\text{ppc}_j$. The total valuation of the matching at this point is $M - (b_{i,j} - \text{ppc}_j) c_j$. But the valuation of the matching at this point also is equal to $M_{-i}$ since lowering the bid below $\text{ppc}_j$ makes the matching no longer assign $i$ to $j$ (and all other edges are deleted, so $i$ is not assigned anywhere else). So we get $M - (b_{i,j} - \text{ppc}_j) c_j = M_{-i}$, and therefore $\text{ppc}_j c_j = M_{-i} - (M - b_{i,j} c_j)$, which is the VCG price.

# References

[1] Gagan Aggarwal. *Privacy Protection and Advertising in a Networked World*. PhD thesis, Stanford University, September 2005.

[2] Gagan Aggarwal, Ashish Goel, and Rajeev Motwani. Truthful auctions for pricing search keywords. In *ACM Conference on Electronic Commerce (EC06)*, June 2006.

[3] E. Clarke. Multipart pricing of public goods. *Public Choice*, 11:17–33, 1971.

[4] B. Edelman, M. Ostrovsky, and M. Schwarz. Internet advertising and the generalized second price auction: Selling billions of dollars worth of keywords. In *Second Workshop on Sponsored Search Auctions*, June 2006.

[5] T. Groves. Incentives in teams. *Econometrica*, 41:617–631, 1973.

[6] A. Mas-Collel, M. Whinston, and J. Green. *Microeconomic Theory*. Oxford University Press, 1995.

[7] Nielsen//NetRatings. Interactive advertising bureau (IAB) search branding study, August 2004. Commissioned by the IAB Search Engine Committee. Available at http://www.iab.net/resources/iab_searchbrand.asp.

[8] Hal Varian. Position auctions, February 2006. Working Paper, available at http://www.sims.berkeley.edu/~hal/Papers/2006/position.pdf.

[9] W. Vickrey. Counterspeculation, auctions and competitive sealed tenders. *Journal of Finance*, 16:8–37, 1961.